\documentclass[psfig,a4paper]{article}

\usepackage{amsmath,latexsym}

\newtheorem{property}{Property}
\def\proof{{\bf{Proof }}} 
\def\fin{\hspace{\stretch{1}}$\Box$}
\def\st{\,:\,}
\newcommand{\eq}[1]{\begin{equation*} #1 \end{equation*}}
\newcommand{\eqa}[1]{\begin{eqnarray*} #1 \end{eqnarray*}}

\def\Tr{{\rm Tr}}
\def\Id{\textbf{1}}

\def\N{\{1\ldots n\}}
\def\bin{\{0,1\}}
\def\bases{\{0,1,2\}}
\def\bell{\{0,1,2,3\}}
\def\w{w}

\def\genK{{\mathcal{G}}}
\def\suppl{{\mathcal{S}}}
\def\scal{\cdot}

\newcommand{\rv}[1]{\boldsymbol{#1}}
\newcommand{\pr}[1]{{\rm P}_{\rv{#1}}}

\newcommand{\ket}[2]{ | \, {#1} \rangle_{#2}}
\newcommand{\bra}[2]{\,_{#2} \langle {#1} \,  |}
\newcommand{\proj}[2]{\ket{#1}{#2}\bra{#1}{#2}}

\newcommand{\ketb}[1]{ | \, {#1} \rangle}
\newcommand{\brab}[1]{ \langle {#1} \,  |}
\newcommand{\projb}[1]{\ketb{#1}\brab{#1}}
\newcommand{\scalarb}[2]{ \langle {#1} \,  | \, {#2}  \rangle}

\def\psim{\Psi^-}

\def\a{\vec a}
\def\b{\vec b}
\def\aa{\vec \alpha}
\def\bb{\vec \beta}
\def\e{\vec e}
\def\c{\vec c}
\def\cc{\vec \gamma}
\def\k{\vec\kappa}
\def\d{\vec d}
\def\ttheta{\vec\theta_{\k}}
\def\dw{d_K}
\def\ens{\scriptscriptstyle}
\def\aaE{\alpha_{\ens E}}
\def\aE{a_{\ens E}}
\def\cE{c_{\ens E}}
\def\aaR{\alpha_{\ens R}}
\def\aR{a_{\ens R}}

\def\cR{c_{\ens R}}
\def\aaT{\alpha_{\ens T}}
\def\aT{a_{\ens T}}
\def\cT{c_{\ens T}}
\def\aaS{\alpha_{\ens S}}
\def\bbS{\beta_{\ens S}}
\def\cS{c_{\ens S}}

\def\coR{c_{\ens \overline{R}}}
\def\ScoR{\begin{array}{l}\scriptstyle \coR\st\\ \scriptstyle \cE\in Y_{\aE},\\ \scriptstyle \cT\in X_{\aT}\end{array}}

\def\P{{\mathcal{P}}}
\def\V{{\mathcal{V}}}
\def\C{{\mathcal{C}}}

\def\X{{\mathcal{X}}}
\def\Y{{\mathcal{Y}}}
\def\phiv{\chi_{v | P}}
\def\A{\Delta}

\def\Uv{U_{v \k \coR}}
\def\Vv{V_{v \k \coR}}
\def\x{\vec x}
\def\y{\vec y}
\def\z{\vec z}
\newcommand{\phivx}[1]{\phi_{v,\coR,\x,#1}}
\newcommand{\psivx}[1]{\psi_{v,\coR,\x,#1}}
\def\pia{\vec\pi_{\aR}}

\begin{document}

\title{Security of EPR-based Quantum Key Distribution using three bases}
\author{Hitoshi Inamori \\ Centre for Quantum Computation, Oxford University}
\date{\today}
\maketitle

\begin{abstract}
Modifications to a previous proof of the security of EPR-based quantum key distribution are proposed. This modified version applies to a protocol using three conjugate measurement bases rather than two. A higher tolerable error rate is obtained for the three-basis protocol.
\end{abstract}

%%%%%%%%%%%%%%%%%%%%%%%%%%%%%%%%%%%%%%%%%%%%%%%%%%%%%%%%%%%%%%%%%%%%%%%%
%		INTRODUCTION
%%%%%%%%%%%%%%%%%%%%%%%%%%%%%%%%%%%%%%%%%%%%%%%%%%%%%%%%%%%%%%%%%%%%%%%%

\section{Introduction}

A modified version of a proof of the security~\cite{Ina00epr2} of EPR-based quantum key distribution is proposed. Based on the works~\cite{Bru98,Lo99}, this modified version applies to a protocol using three conjugate measurement bases rather than two. Only modified parts of the proof are presented here. The framework of our study and the remaining parts of the proof can be found in~\cite{Ina00epr2}.

%%%%%%%%%%%%%%%%%%%%%%%%%%%%%%%%%%%%%%%%%%%%%%%%%%%%%%%%%%%%%%%%%%%%%%%%
%		THE PROTOCOL
%%%%%%%%%%%%%%%%%%%%%%%%%%%%%%%%%%%%%%%%%%%%%%%%%%%%%%%%%%%%%%%%%%%%%%%%

\section{The protocol}

We describe the quantum key distribution protocol under consideration. It is a variation of the protocol described in~\cite{Ina00epr2} using three conjugate measurement bases rather than two. In this variation, the source is supposed to emit pairs of photons with orthogonal polarisations. As a consequence, Alice's bits and Bob's bits are anticorrelated when no error occurs.

\begin{description}
% Protocol Setup
\item[Protocol setup]\hfill

Alice and Bob specify:
\begin{itemize}
\item $m$, the length (in bits) of the private key to be generated.
\item $\epsilon$, the maximum threshold value for the error rate during the quantum transmission ($\epsilon < 1/4$).
\item $\tau$, a security constant such that $\frac{\epsilon}{1-\epsilon} < \frac{\epsilon}{1-\epsilon}+\tau < 1$.
\item the security parameter $r$. It must be large enough so that Alice and Bob can find a binary matrix $K$ of size $m\times r$ such that any linear combination of rows of $K$ that contains at least one row of $K$ has weight greater than $\dw=\left(\frac{\epsilon}{1-\epsilon}+\tau\right) r$. Alice and Bob choose one such matrix $K$. Shannon's coding theorem tells that for asymptotic values of $m$, such matrix can be found if $r$ obeys the inequality:
\eq{
\frac{m}{r} \leq 1 - h\left(\frac{1}{2}\frac{\epsilon}{1-\epsilon}+\frac{\tau}{2}\right).
}
\item An error reconciliation scheme for strings of length $s=\left\lfloor \frac{r}{1-\epsilon} \right\rfloor$ bits as specified in~\cite{Ina00epr2}.
\item $n$, the number of pairs of photons to be sent to the legitimate parties. A good choice for $n$ is $\left\lceil\frac{r}{\frac{1-\epsilon}{3}-\tau_S}\right\rceil$ where $\tau_S$ is a small but strictly positive constant.
\end{itemize}
% Quantum Transmission
\item[Quantum transmission] \hfill
\begin{itemize}
\item A source sends a sequence of $n$ photons to Alice and another sequence of $n$ photons to Bob. It is assumed that ideally, for each $i\in\N$, the source emits a pair of photons in the state:
\eq{
|\psim\rangle = \frac{\ket{0}{0}\ket{1}{0}-\ket{1}{0}\ket{0}{0}}{\sqrt{2}}
}
and that Alice's $i$-th photon is the first photon of this pair, and Bob's $i$-th photon is the second photon of this pair. The kets $\ket{0}{0}$ and $\ket{1}{0}$ form an orthonormal basis ``0'' of the Hilbert space describing the polarisation of one photon. The kets $\ket{0}{1}=\frac{\ket{0}{0}-\ket{1}{0}}{\sqrt{2}}$ and $\ket{1}{1}=\frac{\ket{0}{0}+\ket{1}{0}}{\sqrt{2}}$ form its first conjugate basis ``1''. The second conjugate basis ``2'' is formed by the kets $\ket{0}{2}=\frac{\ket{0}{0}+ i \ket{1}{0}}{\sqrt{2}}$ and  $\ket{1}{2}= i \frac{\ket{0}{0}- i \ket{1}{0}}{\sqrt{2}}$.

As in~\cite{Ina00epr2}, the source needs not to be trusted and can be under Eve's control.
The only assumption is that Alice and Bob receive a sequence of $n$ single photon signals on each side.
\item We assume that the measurement devices of Alice and Bob have efficiency one. For each $i\in\N$,
\begin{enumerate}
\item Alice picks randomly a basis $a_i\in\bases$ with uniform probability distribution. Alice measures her $i$-th photon in the basis $a_i$, obtaining the outcome $\alpha_i\in\bin$, corresponding to the state $\ket{\alpha_i}{{a_i}}$.
\item Similarly, Bob picks randomly and independently of Alice a basis $b_i\in\bases$ with uniform probability distribution. Bob measures his $i$-th photon in the basis $b_i$, obtaining the outcome $\beta_i\in\bin$, corresponding to the state $\ket{\beta_i}{{b_i}}$.
\end{enumerate}
\end{itemize}
% Sifting
\item[Sifting] \hfill

Alice and Bob compare publicly their bases $\a$ and $\b$. We denote by $\d$ the vector in $\bases^n$ defined by $d_i = b_i-a_i \pmod{3}$ for all $i\in\N$. If the number of indexes $i\in\N$ such that $a_i=b_i$ is greater than or equal to $s$ then the sifted set $S$ is the set of the first $s$ such indexes. Otherwise the validation test is failed. The bit strings $\aaS$ and $\neg\bbS$ are the sifted keys, where for any vector $\x\in\bin^n$, we denote by $\neg\x$ the vector whose $i$-th entry is $1+x_i\pmod{2}$ for all $i\in\N$. 

% Error Correction
\item[Error correction]\hfill

Alice and Bob perform the error correction on their sifted keys $\aaS$ and $\neg\bbS$ as specified in the protocol setup.  The error set $E$ is the set of indexes $i$ in $S$ in which an error is found, that is, $\alpha_i = \beta_i$. The error vector $\e$ is the vector in $\bin^s$ giving the positions of the errors ($\forall i\in\{1,\ldots,s\}$, $e_i=1$ if and only if $\alpha_i = \beta_i$).  We denote by $e$ the size of the set $E$. The validation test is passed if $e < \epsilon s$, otherwise it is failed. If the validation test is passed, then the reconciled set $R$ is the set of the first $r$ indexes $i\in S\setminus E$.  Therefore $|R|=r$ and $\forall i\in R$, $a_i=b_i$ and $\alpha_i=\neg\beta_i$. Alice and Bob obtain with high probability an identical string of bits $\aaR\in\bin^r$, called the reconciled key.
% Privacy amplification
\item[Privacy amplification]\hfill

The private key is defined as:
\begin{enumerate}
\item $\k = K \aaR \pmod{2}$ if the validation test is passed.
\item an $m$-bit string $\k$ picked randomly by Alice with uniform probability distribution each time the validation test is failed.
\end{enumerate}
\end{description}

%%%%%%%%%%%%%%%%%%%%%%%%%%%%%%%%%%%%%%%%%%%%%%%%%%%%%%%%%%%%%%%%%%%%%%%%
%		PRIVACY OF THE PROTOCOL
%%%%%%%%%%%%%%%%%%%%%%%%%%%%%%%%%%%%%%%%%%%%%%%%%%%%%%%%%%%%%%%%%%%%%%%%

\section{Privacy of the protocol}

The privacy of the three-basis protocol is stated and the achievable key-creation rate discussed.
\begin{property}
The protocol described above offers perfect privacy: for any eavesdropping strategy chosen by a possible eavesdropper, the conditional entropy of the private key $\rv{\k}$ given the eavesdropper's view $\rv{v}$ is bounded from below by:
\eq{
H(\rv{\k} | \rv{v}) \geq m - 2\left(m+\frac{1}{\ln 2}\right)\left(\theta(r)+2\sqrt{\theta(r)}\right),
}
where
\eq{
\theta(r) = e^{-\frac{1}{16}\tau^3 r}.
}

The above bound applies for any value of the security parameter $r$ such that the matrix $K$ specified in the protocol exists.

\end{property}

Therefore, for asymptotic values of the security parameter $r$, a net gain in shared private bits can be obtained if:
\eq{
1-h\left(\frac{1}{2}\frac{\epsilon}{1-\epsilon}\right)-\frac{1}{1-\epsilon}h(\epsilon) > 0.
}
since $s h(\epsilon)$ bits are required for the one-time pad encryption during the error reconciliation, as in~\cite{Ina00epr2}.

In comparison, we proved that for the original protocol using two measurement bases only, a net gain in shared private bits can be obtained if: 
\eq{
1-h\left(\frac{\epsilon}{1-\epsilon}\right)-\frac{1}{1-\epsilon}h(\epsilon) > 0.
}

Therefore, the three-basis protocol seems to be more robust than the original protocol. However, one must realise that a given threshold value on the error rate gives different physical constraints on the quantum channel depending on whether two-basis or three-basis protocol is used. Whether the two-basis or the three-basis protocol is more robust and efficient depends on the technology chosen to implement the transmission and the reception of the quantum signals.

%%%%%%%%%%%%%%%%%%%%%%%%%%%%%%%%%%%%%%%%%%%%%%%%%%%%%%%%%%%%%%%%%%%%%%%%
%		PROOF OF THE PRIVACY
%%%%%%%%%%%%%%%%%%%%%%%%%%%%%%%%%%%%%%%%%%%%%%%%%%%%%%%%%%%%%%%%%%%%%%%%

\section{Proof of the privacy}

The proof of the above result is given. It is mainly identical to the proof proposed in~\cite{Ina00epr2} for the original protocol. Therefore, only parts that are different from this previous proof are detailed. The main difference between the previous proof and this proof resides in Section \ref{role}.

%%%%%%%%%%%%%%%%%%%%%%%%%%%%%%%%%%%%%%%%%%%%%%%%%%%%%%%%
%	Notations
%%%%%%%%%%%%%%%%%%%%%%%%%%%%%%%%%%%%%%%%%%%%%%%%%%%%%%%%

\subsection{Notations}

We define the notations used throughout the proof. They are similar to the ones used in~\cite{Ina00epr2}, but the indexes for Bell states have been modified. The notations have been adapted to deal with three bases.

\begin{description}

\item[Classical data] \hfill

We denote by $C=(\a,\b,\aa,\bb)$ the classical data Alice and Bob generate during the protocol (after the setup). We denote by $P=(\a,\d,\e)$ the data that are publicly announced by Alice and Bob during the protocol. For any possible $P$, we denote by $\C_P$ the set of values for the classical data that are compatible with the public announcement of $P$. That is, for a given $P=(\a,\d,\e)$, 
\eqa{
\C_P &=& \{ C'=(\a',\b',\aa',\bb') \st \a'=\a,\nonumber\\
& &\forall i,\, b'_i = a_i+d_i\pmod{3}\nonumber\\
& &\forall i\in E,\,\alpha'_i = \beta'_i \mbox{ and } \forall i\in S\setminus E,\,\alpha'_i = \neg \beta'_i \\
& &\mbox{ where $S$ and $E$ are given by $\d$ and $\e$.} \}.
} 

Given a possible $P$ and a value for the private key $\k$, we define $\C_{P,\k}$ as the set of values for the classical data that are compatible with the public announcement of $P$ and generation of $\k$ for the private key. That is, for a given $P=(\a,\d,\e)$, 
\eqa{
\C_{P,\k} &=& \{ C'=(\a',\b',\aa',\bb') \st \a'=\a,\\
& &\forall i,\, b'_i = a_i+d_i\pmod{3}\nonumber\\
& &\forall i\in E,\,\alpha'_i = \beta'_i \mbox{ and } \forall i\in S\setminus E,\,\alpha'_i =\neg \beta'_i \\
& &K \aaR' = \k\pmod{2}, \\
& &\mbox{ where $S$, $E$ and $R$ are given by $\d$ and $\e$.} \}.
} 

Finally, we denote by $\P$ the set of all possible public announcements for which the validation test is passed. That is, 
\eq{
\P = \{ P=(\a,\d,\e) \st \w_0(\d)\geq s \mbox{ and } e < \epsilon s\}.
}

%For any vector $\vec x$ and any symbol $A$, $\w(\vec x)$ is the number of non-zero entries, and $\w_A(\vec x)$ is the number of entries with symbol $A$. 
%Let $X$ be a set, $Y$ be a subset of $X$ and $x$ and $y$ be two elements of $X$. Then we define $\delta_{x, y}$ and $\delta_{x,X}$ by:
%\eqa{\delta_{x,y} &=& \left\{\begin{array}{ll} 1 & \mbox{ if } x=y,\\ 0 & \mbox{ if } x\neq y\end{array}\right.\\ \delta_{x,Y} &=& \left\{\begin{array}{ll} 1 & \mbox{ if } x\in Y,\\ 0 & \mbox{ if } x\notin Y\end{array}\right.}
We denote by $T$ the subset $S\setminus(E\cup R)$, and by $t$ the size of $T$.

\item[Bell states] \hfill

For each $i\in\N$, we define the Bell basis $\{\ket{0}{i},\,\ket{1}{i},\,\ket{2}{i},\,\ket{3}{i}\}$ of the $i$-th pair of photons as:
\eqa{
\ket{0}{i} &=& \frac{\ket{0}{0,i}\ket{1}{0,i}-\ket{1}{0,i}\ket{0}{0,i}}{\sqrt{2}},\\
\ket{1}{i} &=& \frac{\ket{0}{0,i}\ket{1}{0,i}+\ket{1}{0,i}\ket{0}{0,i}}{\sqrt{2}},\\
\ket{2}{i} &=& \frac{\ket{0}{0,i}\ket{0}{0,i}-\ket{1}{0,i}\ket{1}{0,i}}{\sqrt{2}},\\
\ket{3}{i} &=& \frac{\ket{0}{0,i}\ket{0}{0,i}+\ket{1}{0,i}\ket{1}{0,i}}{\sqrt{2}},
}
where the first and the second state in the product states in the rhs.~correspond to Alice's and Bob's $i$-th photon's polarisation state, respectively. 
%Tensor products are implied when we consider state of several photons, that is, $\ket{\aa,\bb}{\a,\b}=\otimes_{i=1}^n \ket{\alpha_i}{a_i,i}\ket{\beta_i}{b_i,i}$ and $\ketb{\c}=\otimes_{i=1}^n \ket{c_i}{i}$. For any subset $X$ of $\N$, $\ket{\alpha_{\ens X},\beta_{\ens X}}{a_{\ens X},b_{\ens X}}=\otimes_{i\in X} \ket{\alpha_i}{a_i,i}\ket{\beta_i}{b_i,i}$.

Given a basis $a\in\bases$, we define $X_a$ as the set of indexes of Bell states that are compatible with Alice and Bob measuring in the same basis $a$ and obtaining opposite bit values (corresponding to a faithful transmission, as the source is supposed to emit an antisymmetric state). Likewise, we define $Y_a$ as the set of indexes of Bell states that are compatible with Alice and Bob measuring in basis $a$ and sharing the same bit value (corresponding to an error). That is, $X_0 = \{0,1\}$, $X_1 = \{0,2\}$, $X_2 = \{0,3\}$, $Y_0 = \{2,3\}$, $Y_1 = \{1,3\}$ and $Y_2=\{1,2\}$. Given the choice of bases $\a$ and a set $A\subset\N$, we define $X_{a_{\ens A}}$ as $\{ c_{\ens A}\in\bell^A \st \forall i\in A, c_i\in X_{a_i}\}$ and $Y_{a_{\ens A}}$ as $\{ c_{\ens A}\in\bell^A \st \forall i\in A, c_i\in Y_{a_i}\}$. Given a reconciled set $R$ and the choice of bases $\aR$ on $R$, for any $\cR\in X_{\aR}$, we will denote by $\cc$ the unique $\cc\in\bin^r$ such that for each $i\in\{1,\ldots,r\}$, $c_i=(1+a_i)\gamma_i$. For any vectors $\vec x$, $\vec y$ $\in\bin^r$, we define $\x \scal\y$ as $\vec x \scal \vec y \stackrel{Def}{=} \sum_{i=1}^r x_i y_i$. Given $R$ and $\aR$,  for any $\cR\in X_{\aR}$, we have the identity $\bra{\aaR,\neg\aaR}{\aR} \cR \rangle = \frac{(-1)^{\aaR\scal(\neg\cc)}(-i)^{\pia\scal\cc}}{\sqrt{2}^r}$, where $\pia$ is a vector in $\bin^r$ with its $i$-th entry equal to 1 if and only if $a_i=2$. 

\end{description}

%%%%%%%%%%%%%%%%%%%%%%%%%%%%%%%%%%%%%%%%%%%%%%%%%%%%%%%%
%	Model of measurements
%%%%%%%%%%%%%%%%%%%%%%%%%%%%%%%%%%%%%%%%%%%%%%%%%%%%%%%%

\subsection{Model of measurements}

The mathematical model of measurements on the quantum state generated by the source is almost identical to the one described in~\cite{Ina00epr2}. 

The state of the $n$ couples of photons and the probe created by Eve reads as:
\eq{
\rho = \sum_{\c,\c'}\ketb{E_{\c}}\brab{E_{\c'}}\otimes\ketb{\c}\brab{\c'},
}
where the states $\ketb{E_{\c}}$ are states of Eve's probe that are possibly nor orthogonal nor normalised.
The positive operator giving the probability that Alice and Bob get $C=(\a,\b,\aa,\bb)$ as their classical data is:
\eq{
F_C=\pr{\a}(\a)\pr{\b}(\b)\proj{\aa,\bb}{\a,\b},
}
where $\pr{\a}(\a)=1/3^n$ and $\pr{\b}(\b)=1/3^n$ for any choice of $\a$ and $\b$. Note that since for all $i\in\N$, $d_i=b_i-a_i\pmod{3}$, we have $\pr{\a}(\a)\pr{\b}(\b)=\pr{\a}(\a)\pr{\d}(\d)$ where $\pr{\d}(\d)=1/3^n$.

The positive operator giving the probability that Alice and Bob publicly announce $P=(\a,\d,\e)$ while they get the private key $\k$ is:
\eqa{
F_{P,\k}
&=&\pr{\a}(\a)\pr{\d}(\d)\Id_{\overline{S}}\otimes\sum_{\aaE\in\bin^e}\proj{\aaE,\aaE}{\aE,\aE}\nonumber\\
& &\otimes \sum_{\aaT\in\bin^t}\proj{\aaT,\neg\aaT}{\aT,\aT}\nonumber\\
& &\otimes\sum_{\begin{array}{l}\scriptstyle \aaR\in\bin^r\st\\ \scriptstyle K\aaR = \k \pmod{2}\end{array}}\proj{\aaR,\neg\aaR}{\aR,\aR},
%&=& \pr{\a}(\a)\pr{\d}(\d)\Id_{\overline{S}}\otimes\sum_{\begin{array}{l}\scriptstyle \aaS\in\bin^s\st\\ \scriptstyle K\aaS = \k\end{array}}\proj{\aaS,\aaS+\e}{\aS,\aS}
}
and the positive operator giving the marginal probability that Alice and Bob publicly announce $P=(\a,\d,\e)$ is:
\eqa{
F_{P}
&=&\pr{\a}(\a)\pr{\d}(\d)\Id_{\overline{S}}\otimes\sum_{\aaE\in\bin^e}\proj{\aaE,\aaE}{\aE,\aE}\nonumber\\
& &\otimes \sum_{\aaT\in\bin^t}\proj{\aaT,\neg\aaT}{\aT,\aT}\nonumber\\
& &\otimes\sum_{\aaR\in\bin^r}\proj{\aaR,\neg \aaR}{\aR,\aR}.
%&=&\pr{\a}(\a)\pr{\d}(\d)\Id_{\overline{S}}\otimes\sum_{\aaS\in\bin^s}\proj{\aaS,\aaS}{\aS,\aS}
}

Note again that when no error occurs, Alice's bit and Bob's bit are anticorrelated.

We denote by $\V_{P}$ the set of views $v$ that are compatible with the public announcement $P$. The positive operator giving the probability that Eve gets the view $v$ given that Alice and Bob announced $P$ will be denoted by $G_{v | P}=\projb{\phiv}$, where we assume again without loss of generality that the operators $G_{v | P}$ are of rank one.

%%%%%%%%%%%%%%%%%%%%%%%%%%%%%%%%%%%%%%%%%%%%%%%%%%%%%%%%
%	The role of the validation test
%%%%%%%%%%%%%%%%%%%%%%%%%%%%%%%%%%%%%%%%%%%%%%%%%%%%%%%%

\subsection{The r\^ole of the validation test}\label{role}

Here a variation of the Property 2 in~\cite{Ina00epr2} is given. It is shown that when three bases are used for the validation test, the constraint on the photon state created by Eve is more stringent.  
Given a possible reconciled set $R$, let $\Pi_R$ be the orthogonal projection operator defined as:
\eqa{
\Pi_R &=& \sum_{\begin{array}{l} \scriptstyle\c\in\bell^n\st \\ \scriptstyle\w(\cR) \geq \dw/2\end{array}} \projb{\c}\\
&=& \Id_{\overline{R}} \otimes \sum_{\begin{array}{l}\scriptstyle\cR\in\bell^r\st\\ \scriptstyle\w(\cR)\geq \dw/2 \end{array}}\projb{\cR}.
} 

The following property is then proved.

\begin{property}
The eigenvalues of the semi-definite positive Hermitian operator
\eq{
\sum_{P\in \P} \Pi_R F_P \Pi_R,
}
where $R$ is specified by $P$ in the sum, are bounded from above by
\eq{
\theta(r) = e^{-\frac{1}{16}\tau^3 r}.
}
\end{property}

\proof 
The above operator can be written as:
\eq{
\sum_{P\in \P} \Pi_R F_P \Pi_R = \sum_{\begin{array}{l}\scriptstyle\d\in\bin^n\st\\ \scriptstyle \w_0(\d) \geq s\end{array}}\sum_{\begin{array}{l}\scriptstyle\e\in\bin^s\st\\ \scriptstyle \w(\e) <\epsilon s\end{array}} \Pi_R \left(\sum_{\a} F_P\right)\Pi_R.
}
Now for given $\d$ and $\e$,
\eq{
\sum_{\a} F_P = \pr{\d}(\d) \Id_{\overline{S}} \otimes_{i\in E} Y_i \otimes_{j\in T} X_j \otimes_{k\in R} X_k,
}
where
\eqa{ 
X_i &=&  \projb{0}+\frac{1}{3}\projb{1}+\frac{1}{3}\projb{2}+\frac{1}{3}\projb{3}, \\
Y_i &=&  \frac{2}{3}\projb{1}+\frac{2}{3}\projb{2}+\frac{2}{3}\projb{3}
}
are operators acting on $i$-th photon pair's Hilbert space. The last equalities are derived directly from the definition of the Bell states. As a consequence, we have,
\eq{
\Pi_R \left(\sum_{\a} F_P\right) \Pi_R = \pr{\d}(\d) \Id_{\overline{S}}\otimes_{i\in E} Y_i \otimes_{j\in T}X_j\otimes\Big(\sum_{\begin{array}{l} \scriptstyle \cR \in\bell \st\\ \scriptstyle \w(\cR)\geq \dw/2\end{array}} \frac{\projb{\cR}}{3^{\w(\cR)}}\Big). 
}

Now, given $\d\in\bin^n$, the operator:
\eq{
\sum_{\e\st \w(\e)<\epsilon s} \Pi_R \left(\sum_{\a} F_P\right) \Pi_R
}
is diagonal in the Bell basis $\ketb{\c}$. Given a vector $\c\in\bell^n$ and an error vector $\e\in\bin^s$, a necessary condition for the scalar:
\eq{
\brab{\c}\Pi_R \left(\sum_{\a} F_P\right) \Pi_R\ketb{\c}
}
to be non zero is that for all $i\in S$, $e_i=0$ if $c_i=0$ and $\w(\cS)\geq \frac{\dw}{2}+e$. There are $\binom{\w(\cS)}{e}$ such vectors $\e$ of weight $e$, if $0\leq e <\epsilon s$ and $e\leq \w(\cS)-\dw/2$. Therefore,
\eqa{
\lefteqn{\brab{\c}\sum_{\e\st \w(\e)<\epsilon s}\Pi_R \left(\sum_{\a} F_P\right) \Pi_R\ketb{\c}}\nonumber\\
&\leq& \pr{\d}(\d) \sum_{\begin{array}{l}\scriptstyle 0\leq e < \epsilon s\\ \scriptstyle e \leq \w(\cS)-\dw/2\end{array}} \binom{\w(\cS)}{e}\left(\frac{2}{3}\right)^e\left(\frac{1}{3}\right)^{\w(\cS)-e}.
}

Now, $\dw$ is either greater or smaller than $\frac{2}{3}\w(\cS)\left(1+\tau(1-\epsilon)\right)$.
\begin{itemize}
\item If $\dw > \frac{2}{3}\w(\cS)\left(1+\tau(1-\epsilon)\right)$, then
\eq{
\w(\cS)-\frac{\dw}{2} < \frac{2}{3}\w(\cS)\left(1-\frac{1}{2}\tau(1-\epsilon)\right) \,\mbox{ and,}
}
\item if $\dw \leq \frac{2}{3}\w(\cS)\left(1+\tau(1-\epsilon)\right)$, then
\eqa{
\epsilon s&\leq& \frac{\epsilon r}{1-\epsilon}\\
&=& \dw-\tau r\\
&\leq& \frac{2}{3}\w(\cS)\left(1-\frac{1}{2}\tau(1-\epsilon)\right),
}
where we have used $r\geq s(1-\epsilon)$ and $s\geq \w(\cS)$.
\end{itemize}
We thus derived that:
\eqa{
\lefteqn{\brab{\c}\sum_{\e\st \w(\e)<\epsilon s}\Pi_R \left(\sum_{\a} F_P\right) \Pi_R\ketb{\c}}\nonumber\\
&\leq& \pr{\d}(\d) \sum_{0\leq e <\frac{2}{3}\w(\cS)\left(1-\frac{1}{2}\tau(1-\epsilon)\right) } \binom{\w(\cS)}{e}\left(\frac{2}{3}\right)^e\left(\frac{1}{3}\right)^{\w(\cS)-e}\\
&\leq& \pr{\d}(\d) e^{-\frac{2}{9} \tau^2(1-\epsilon)^2\w(\cS)}\\
&\leq& \pr{\d}(\d) e^{-\frac{1}{16}\tau^3 r}\\
&=& \pr{\d}(\d) \theta(r)
}
where we have used the binomial inequality stating that $\sum_{0\leq k < (p-t) n} \binom{n}{k} p^k(1-p)^{n-k} \leq e^{-2 t^2 n}$ for any positive integer $n$ and $0 < p-t \leq p <1$. In the last inequality we have used the inequalities $\epsilon <1/4$ and $\w(\cS)\geq\dw/2$ when the above scalar is non zero.

Remarking that the operator $\sum_{\a,\e\st\ e<\epsilon s}\Pi_R F_P\Pi_R$ is diagonal in the Bell basis for all $\d$ and $\sum_{\d \st \w_0(\d) \geq s} \pr{\d}(\d) \leq 1$, this concludes the proof.\fin

The above property implies that:
\eq{
\Tr\Big(\Id_{\mbox{\scriptsize Eve}}\otimes\sum_{P\in\P} \Pi_R F_P \Pi_R \rho\Big) \leq \theta(r)
}
where $\Id_{\mbox{\scriptsize Eve}}$ is the identity operator acting on the Hilbert space of the probe. That is,
\eq{
\sum_{P\in\P}\pr{\a}(\a)\pr{\d}(\d) \sum_{\begin{array}{l}\scriptstyle \coR \st \\\scriptstyle \cE\in Y_{\aE},\\\scriptstyle\cT\in X_{\aT}\end{array}}\sum_{\begin{array}{l}\scriptstyle\cR\in X_{\aR}\st\\\scriptstyle\w(\cR) \geq \dw/2\end{array}} \scalarb{E_{\c}}{E_{\c}} \leq \theta(r).
}

%%%%%%%%%%%%%%%%%%%%%%%%%%%%%%%%%%%%%%%%%%%%%%%%%%%%%%%%
%	Quasi-independence of the key and the view
%%%%%%%%%%%%%%%%%%%%%%%%%%%%%%%%%%%%%%%%%%%%%%%%%%%%%%%%

\subsection{Quasi-independence of the key and the view}

In this section we compute the joint probability distribution of the key and the view. We prove that this distribution is very close to a product of an uniform distribution for the key and the marginal probability distribution of the view. This section is identical to the Section 5.4 of the previous proof, except for the apparition of few phase factors that do not appear in the final result.

\begin{property}
For any given eavesdropping strategy chosen by Eve and returning a view $\rv{v}$, the probability distribution of the key $\rv{\k}$ and the view $\rv{v}$ obeys the following inequality:
\eq{
\sum_{P\in\P}\sum_{v\in\V_P}\sum_{\k\in\bin^m} \left|\pr{\k v}(\k, v)-\frac{1}{2^m}\pr{v}(v)\right | \leq 2\left(\theta(r)+2\sqrt{\theta(r)}\right)
}
where $m$ is the length of the private key and $r$ is the size of the reconciled set.
\end{property}

\proof For any $\k\in\bin^m$, $P$ and $v\in\V_P$, we have:
\eqa{
\lefteqn{\pr{\k v}(\k,v) - \frac{1}{2^m}\pr{v}(v)}\nonumber\\
&=& \Tr(G_{v | P}\otimes F_{P,\k}\, \rho) - \frac{1}{2^m} \Tr(G_{v|P}\otimes F_P\,\rho) \\
%&=& \pr{\a}(\a)\pr{\d}(\d)\sum_{\c,\c'} \brab{E_{\c'}} G_{v|P} \ketb{E_{\c}} \delta_{\coS,\coS'}\nonumber\\& & \times \brab{\cE'}\Big(\sum_{\aaE\in\bin^e}\proj{\aaE,\aaE}{\aE,\aE}\Big)\ketb{\cE}\nonumber\\& & \times \brab{\cT'}\Big(\sum_{\aaT\in\bin^t}\proj{\aaT,\neg\aaT}{\aT,\aT}\Big)\ketb{\cT}\nonumber\\& & \times \brab{\cR'}\Big[\sum_{\begin{array}{l}\scriptstyle\aaR\in\bin^r\st\\\scriptstyle K\aaR=\k\end{array}}\proj{\aaR,\neg\aaR}{\aR,\aR}\nonumber\\& & -\frac{1}{2^m}\sum_{\aaR\in\bin^r}\proj{\aaR,\neg\aaR}{\aR,\aR}\Big]\ketb{\cR}\\
&=& \pr{\a}(\a)\pr{\d}(\d)\sum_{\begin{array}{l}\scriptstyle \c,\c'\st\\ \scriptstyle \cE,\cE'\in Y_{\aE},\\ \scriptstyle \cT,\cT'\in X_{\aT},\\ \scriptstyle \cR,\cR'\in X_{\aR}\end{array}} \brab{E_{\c'}} G_{v|P} \ketb{E_{\c}}\delta_{\coR,\coR'} d_{\k, \aR}(\cc,\cc'),
}
where
\eq{
d_{\k,\aR}(\cc,\cc') =  (-i)^{\pia\cdot\cc}(+i)^{\pia\cdot\cc'}\sum_{\begin{array}{l}\scriptstyle \aaR\in\bin^r\st\\ \scriptstyle K\aaR = \k\pmod{2}\end{array}} \frac{(-1)^{\aaR \scal (\cc+\cc')}}{2^r} - \frac{1}{2^m}\delta_{\cc,\cc'}.
}
where we have used the identity $\bra{\aaR,\neg\aaR}{\aR,\aR} \cR\rangle = \frac{(-1)^{\aaR\scal(\neg\cc)}(-i)^{\pia\cdot\cc}}{\sqrt{2}^r}$ for any $\cR\in X_{\aR}$ (note that $\neg\cc+\neg\cc'=\cc+\cc'$). 

Let's define $\genK$ as the set of all linear combinations over $\bin$ of rows of $K$. It has been proved in~\cite{Ina00epr2} that:
\eq{
\sum_{\begin{array}{l}\scriptstyle \aaR\in\bin^r\st\\ \scriptstyle K\aaR = \k\pmod{2}\end{array}} (-1)^{\aaR \scal (\cc+\cc')}  = \left\{ \begin{array}{ll} (-1)^{\ttheta\scal(\cc+\cc')}2^{r-m} & \,\mbox{ if }\, \cc+\cc' \in \genK, \\ 0 & \,\mbox{ if }\, \cc+\cc'\notin\genK \end{array}\right. 
}
where $\ttheta$ is a vector in $\bin^r$ such that $K\ttheta = \k \pmod{2}$. We have:
\eq{
\pr{\k v}(\k,v) - \frac{1}{2^m}\pr{v}(v) = \frac{1}{2^m} \pr{\a}(\a)\pr{\d}(\d) \sum_{\begin{array}{l}\scriptstyle \coR\st\\ \scriptstyle \cE\in Y_{\aE},\\\scriptstyle\cT\in X_{\aT}\end{array}} (\Uv+\Vv)^\dag \A (\Uv+\Vv),
}
where $\Uv$ and $\Vv$ are complex vectors of dimension $2^r$ and $\A$ is a $2^r\times 2^r$ complex matrix, whose entries are indexed by $\cc\in\bin^r$. The $\cc$-th entry of $\Uv$ and $\Vv$ are:
\eqa{
\Big( \Uv \Big)_{\cc} &=& \left\{\begin{array}{cl}(-1)^{\ttheta\scal\cc}(-i)^{\pia\cdot\cc}\brab{\phiv} E_{\c} \rangle & \,\mbox{ if }\, \w(\cc) < \dw / 2, \\ 0 &\,\mbox{ if }\, \w(\cc) \geq \dw/2. \end{array}\right.\\
\Big( \Vv \Big)_{\cc} &=& \left\{\begin{array}{cl}0 &\,\mbox{ if }\, \w(\cc) < \dw/2,\\(-1)^{\ttheta\scal\cc}(-i)^{\pia\cdot\cc}\brab{\phiv} E_{\c} \rangle & \,\mbox{ if }\, \w(\cc) \geq \dw / 2, \\ \end{array}\right.
}
where $\c$ is given by $\coR$ and $\cc$. The $(\cc,\cc')$-th entry of $\A$ is, as in~\cite{Ina00epr2}:
\eq{
\Big( \A \Big)_{\cc, \cc'} = \left\{ \begin{array}{ll} 1 &\,\mbox{ if }\, \cc+\cc'\in\genK\setminus \{0\},\\ 0 &\,\mbox{ if }\, \cc+\cc'\notin\genK\setminus \{0\}.\end{array}\right.
}

This implies $\Uv^\dag\A\Uv=0$, since $\w(\cc) <\dw/2$ and $\w(\cc') < \dw/2$ implies that $\w(\cc+\cc') < \dw$, that is, $\cc+\cc'\notin\genK\setminus\{\vec 0\}$. The matrix $\A$ is diagonalised in the same manner as in~\cite{Ina00epr2}, and we obtain:
\eqa{
\Vv^\dag \A\Vv &=& 2^m \Big[ (2^m-1)  \sum_{\x\in\suppl} |\psivx{\k}|^2-\sum_{\begin{array}{l}\scriptstyle \x\in\suppl \\ \scriptstyle \vec \sigma \in\bin^m\setminus{\vec 0} \end{array}} |\psivx{\k+\vec\sigma}|^2 \Big],\\
\Uv^\dag \A\Vv &=& 2^m \Big[ (2^m-1)  \sum_{\x\in\suppl} |\phivx{\k}^*\psivx{\k}|- \sum_{\begin{array}{l}\scriptstyle \x\in\suppl \\ \scriptstyle \vec \sigma\in\bin^m\setminus{\vec 0} \end{array}}|\phivx{\k+\vec\sigma}^*\psivx{\k+\vec\sigma}|\Big].
}
where $\suppl$ is a subspace of $\bin^r$ supplement to $\genK$, and for any $\z\in\bin^m$,
\eqa{
\phivx{\z} &=& \sum_{\begin{array}{l}\scriptstyle\y\in\genK\st\\\scriptstyle \w(\x+\y) < \dw/2\end{array}} (-i)^{\pia\cdot \y}  \frac{(-1)^{\vec\omega_{\y}\scal\z}}{2^m}\scalarb{\phiv}{E_{\c}},\\
\psivx{\z} &=& \sum_{\begin{array}{l}\scriptstyle\y\in\genK\st\\\scriptstyle \w(\x+\y) \geq \dw/2\end{array}}(-i)^{\pia\cdot \y}\frac{(-1)^{\vec\omega_{\y}\scal\z}}{2^m}\scalarb{\phiv}{E_{\c}}.
}
where for any $\vec y \in \genK$, $\vec\omega_{\vec y}$ is the unique vector in $\bin^m$ such that $K^T\vec\omega_{\vec y}=\vec y\pmod{2}$.

From this one derives the inequality:
\eq{
\sum_{P\in\P}\sum_{v\in\V_P}\sum_{\k\in\bin^m}\left|\pr{\k v}(\k,v) - \frac{1}{2^m}\pr{v}(v)\right|
\leq 2(2^m-1)(\eta + 2\sqrt{\eta}\sqrt{\xi}),
}
where
\eqa{
\eta &=& \sum_{P\in\P}\sum_{\ScoR}\sum_{v\in\V_P}\sum_{\x\in\suppl}\sum_{\k}  \pr{\a}(\a)\pr{\d}(\d) |\psivx{\k}|^2,\\
\xi  &=& \sum_{P\in\P}\sum_{\ScoR}\sum_{v\in\V_P}\sum_{\x\in\suppl}\sum_{\k}  \pr{\a}(\a)\pr{\d}(\d) |\phivx{\k}|^2.
}

We then derive an upper-bound on $\eta$ and $\xi$. We have:
\eqa{
\eta &=&  \sum_{P\in\P}\pr{\a}(\a)\pr{\d}(\d)\sum_{\ScoR}\sum_{v\in\V_P}\sum_{\x\in\suppl}\sum_{\begin{array}{l}\scriptstyle \y, \y'\in\genK \st\\ \scriptstyle \w(\x+\y) \geq \dw/2 \\ \scriptstyle \w(\x+\y') \geq \dw/2\end{array}}\sum_{\k} \frac{(-1)^{\vec\omega_{\y+\y'}\scal\k}}{2^{2m}}\nonumber\\
& &\times (-i)^{\pia\cdot\y}(+i)^{\pia\cdot\y'}\scalarb{E_{\c'}}{\phiv}\scalarb{\phiv}{E_{\c}}\\
&=& \frac{1}{2^{m}}\sum_{P\in\P}\pr{\a}(\a)\pr{\d}(\d)\sum_{\ScoR}\sum_{\x\in\suppl}\sum_{\begin{array}{l}\scriptstyle\y\in\genK \st\\\scriptstyle \w(\x+\y)\geq \dw/2\end{array}}\sum_{v\in\V_P} \scalarb{E_{\c}}{\phiv}\scalarb{\phiv}{E_{\c}}\\
%&=& \frac{1}{2^{m}}\sum_{P\in\P}\pr{\a}(\a)\pr{\d}(\d)\sum_{\ScoR}\sum_{\begin{array}{l}\scriptstyle\cR\in X_{\aR}\st\\\scriptstyle \w(\cR) \geq \dw/2 \end{array}} \scalarb{E_{\c}}{E_{\c}}\\
&\leq& \frac{1}{2^{m}} \theta(r),
}
using the result of the previous section. Similarly,
\eq{
\xi 
%&=& \frac{1}{2^{m}}\sum_{P\in\P}\pr{\a}(\a)\pr{\d}(\d)\sum_{\ScoR}\sum_{\begin{array}{l}\scriptstyle\cR\in X_{\aR}\st\\\scriptstyle \w(\cR) < \dw/2 \end{array}} \scalarb{E_{\c}}{E_{\c}}\\
\leq \frac{1}{2^{m}}.
}

Consequently,
\eqa{
\lefteqn{\sum_{P\in\P}\sum_{v\in\V_P}\sum_{\k\in\bin^m}\left|\pr{\k v}(\k,v) - \frac{1}{2^m}\pr{v}(v)\right|}\nonumber\\
&\leq&  2\left(\theta(r) + 2\sqrt{\theta(r)}\right)
}
which concludes our proof.\fin

%%%%%%%%%%%%%%%%%%%%%%%%%%%%%%%%%%%%%%%%%%%%%%%%%%%%%%%%
%	Bound on the conditional entropy
%%%%%%%%%%%%%%%%%%%%%%%%%%%%%%%%%%%%%%%%%%%%%%%%%%%%%%%%

\subsection{Bound on the conditional entropy}

As in~\cite{Ina00epr2}, we conclude the proof of privacy thanks to the following property.

\begin{property}
Let $\rv{x}$ and $\rv{y}$ be two discrete random variables taking values in the sets $\X$ and $\Y$ respectively. Let $\mu$ be a nonnegative real number. If the following inequality is satisfied:
\eq{
\sum_{x\in\X,\,y\in\Y} \left| \pr{x y}(x,y)-\frac{1}{|\X|}\pr{y}(y) \right| \leq \mu,
}
then the conditional entropy of $\rv{x}$ given $\rv{y}$ is lower-bounded by:
\eq{
H(\rv{x} | \rv{y}) \geq (1-\mu)\log_2 |\X| -\frac{1}{\ln 2} \mu.
}
\end{property}

The proof of this property has been given in~\cite{Ina00epr2}. The probability distribution of the private key and the view obeys the inequality:
\eq{
\sum_{\begin{array}{l}\scriptstyle\k\in\bin^m,\\\scriptstyle v\in\V\end{array}} \left| \pr{\k v}(\k,v)-\frac{1}{2^m} \pr{v}(v) \right| \leq 2(\theta(r)+2\sqrt{\theta(r)}),
}
where we have used the fact that the key is randomly chosen by Alice with uniform probability distribution if the validation test is not passed. Applying the above property for the random variables $\rv{\k}$ and $\rv{v}$, we obtain:
\eq{
H(\rv{\k} | \rv{v}) \geq m-2\left(m+\frac{1}{\ln 2}\right)\left(\theta(r)+2\sqrt{\theta(r)}\right),
}
which concludes the proof of privacy.\fin

%%%%%%%%%%%%%%%%%%%%%%%%%%%%%%%%%%%%%%%%%%%%%%%%%%%%%%%%%%%%%%%%%%%%%%%%%%
%		Acknowledgement						 %
%%%%%%%%%%%%%%%%%%%%%%%%%%%%%%%%%%%%%%%%%%%%%%%%%%%%%%%%%%%%%%%%%%%%%%%%%%

\smallskip
{\bf{Acknowledgements }} The author gratefully acknowledges support provided by the European TMR Network ERP-4061PL95-1412, and thanks Hans Briegel, Artur Ekert, Nicolas Gisin, Patrick Hayden, Norbert L\"utkenhaus, Dominic Mayers, Michele Mosca, Luke Rallan, Peter Shor and Vlatko Vedral for interesting discussions and helpful comments. 

%{\bf{Acknowledgement }} The author gratefully acknowledges....

%%%%%%%%%%%%%%%%%%%%%%%%%%%%%%%%%%%%%%%%%%%%%%%%%%%%%%%%%%%%%%%%%%%%%%%%%%
%		Bibliography						 %
%%%%%%%%%%%%%%%%%%%%%%%%%%%%%%%%%%%%%%%%%%%%%%%%%%%%%%%%%%%%%%%%%%%%%%%%%%

%\bibliographystyle{unsrt}
%\bibliography{../bibliQKD}

\begin{thebibliography}{1}

\bibitem{Ina00epr2}
H.~Inamori.
\newblock Security of {EPR}-based quantum key distribution.
\newblock quant-ph/0008064, 2000.

\bibitem{Bru98}
D.~Bru\ss.
\newblock Optimal eavesdropping in quantum cryptography with six states.
\newblock {\em Phys.~Rev.~Lett.}, 81:3018, 1998.

\bibitem{Lo99}
H.-K. Lo.
\newblock A simple proof of the unconditional security of quantum key
  distribution.
\newblock quant-ph/9904091, 1999.

\end{thebibliography}

%%%%%%%%%%%%%%%%%%%%%%%%%%%%%%%%%%%%%%%%%%%%%%%%%%%%%%%%%%%%%%%%%%%%%%%%%%
%		Appendix						 %
%%%%%%%%%%%%%%%%%%%%%%%%%%%%%%%%%%%%%%%%%%%%%%%%%%%%%%%%%%%%%%%%%%%%%%%%%%

\end{document}